\documentclass{article}

\begin{document}

\newcommand{\tbox}[1]{\mbox{\tiny #1}}
\newcommand{\half}{\mbox{\small $\frac{1}{2}$}}
\newcommand{\mbf}[1]{{\mathbf #1}}


\title{
\mbox{Quantum Chaos, Irreversibility, dissipation and dephasing}
}

\author{Doron Cohen \\
Department of Physics, Ben-Gurion University, Beer-Sheva, Israel.
}

\date{}

\maketitle


\abstract{
The main idea of "Quantum Chaos" studies
is that Quantum Mechanics introduces two
energy scales into the study of chaotic systems:
One is obviously the mean level spacing $\Delta\propto\hbar^d$,
where $d$ is the dimensionality; The other is
$\Delta_b\propto\hbar$, which is known as the
non-universal energy scale, or as the bandwidth,
or as the Thouless energy. Associated with these two
energy scales are two special quantum-mechanical (QM) regimes
in the theory of driven system. These are the
QM adiabatic regime, and the QM non-perturbative regime
respectively. Otherwise Fermi golden rule applies,
and linear response theory can be trusted.
Demonstrations of this general idea, that had been published in 1999,
have appeared in studies of wavepacket dynamics, survival probability,
dissipation, quantum irreversibility, fidelity and dephasing.
The following presentation is intended for non-specialists.
}


\ \\

Driven quantum systems, described by Hamiltonian ${\cal H}(Q,P,x(t))$,
\linebreak  where~$x(t)$ is a time dependent parameter,
are of interest in mesoscopic physics (quantum dots),
as well as in nuclear, atomic and molecular physics.
Due to the time dependence of $x(t)$, the energy of
the system is not a constant of motion.
Rather the system makes "transitions" between energy levels.
Notions such as "survival probability" and "dissipation",
just emphasize particular aspects of the energy spreading process. \\

The name "Quantum Mechanics" is associated with the idea
that the energy is quantized. For generic (chaotic) system
the mean level spacing is $\Delta\propto\hbar^d$,
where $d$ is the dimensionality of the system.
However, one should recognize that there is a second
energy scale $\Delta_b\propto\hbar$ which is introduced
by Quantum Mechanics. This $\hbar$ energy scale is
known in the literature as the "non-universal" energy scale (Berry),
or as the "bandwidth" (Feingold and Peres) or as the Thouless energy. \\

The main focus of Quantum chaos studies (so far) was on
issues of spectral statistics. In this context it turns
out that the sub-$\hbar$ statistical features of the
energy spectrum are "universal", and obey the predictions
of random matrix theory. Non universal (system specific)
features are reflected only in the large scale properties
of the spectrum (hence we can "hear the shape of the drum").
This is the reason why $\Delta_b$ is know as the "non-universal"
energy scale. \\

Having two quantal energy scales implies the existence
of two special QM regimes in the theory of driven systems.
The simplest demonstration of this idea is in the context
of linear driving ($\dot{x}=V$).
The existence of the QM adiabatic regime (very very small $V$)
is associated with having finite $\Delta$.
The existence of the QM non-perturbative regime
(where $V$ is quantum mechanically large,
but still classically small) is associated with
the energy scale $\Delta_b$. \\

{\em Heuristic explanation of having QM adiabatic regime}:
If the rate $V$ is very very small, the system remains all
the time in the same level. This is because there are QM recurrences
that block the attempt to make a transition to any other level.
Obviously this effect is related to having
finite (non-zero) level spacing. \\

{\em Heuristic explanation of having QM non-perturbative regime}:
If $V$ is large enough to induce transitions between levels,
then we have to ask what is the maximal size of a single "step"
in energy space. (Here "step" means first order transition).
The answer is that the maximal step is the bandwidth $\Delta_b$.
The energy spreading process after many "steps" becomes diffusive.
If $V$ is too large, then the breakdown of perturbation theory happens
before even a single step is taken. In such case perturbation theory
cannot be used in order to analyze the energy spreading process. \\

The identification of the "non-perturbative" regime is the
main observation of \mbox{[Cohen PRL 1999]}. An associated
observation is that the semiclassical limit is contained
in the non-perturbative regime. The idea has been applied
and generalized to the analysis of the energy spreading
process (the decay of the survival probability, and the
growth of the variance) in case of "wavepacket dynamics"
[main collaborator: Kottos]. In these studies the control
is over the strength/amplitude $A$ off the perturbation.
The idea also has been generalized to the case of periodically
driven systems \mbox{[Cohen and Kottos, PRL 2000]}, where the control
is over both the rate of change $V$
and the amplitude $A$ of the driving. \\

Recently the idea of having a non-perturbative/semiclassical
regime has been adopted \mbox{[Jacquod, Silvestrov and Beenakker PRE 2001]}
into the context of quantum-irreversibility
studies (also known as "fidelity" or "Loschmidt echo" studies).
Also here, if the perturbation strength $A$ is large enough
("large" in a quantum mechanical sense, but still assumed to be
small in a classical sense), one gets into a semiclassical regime.
This is the same idea as in our "survival probability" studies
in the context of "wavepacket dynamics". But in this particular
context one can further argue \mbox{[Jalabert and Pastawski, PRL 2001]}
that the "semiclassical decay" is in fact
a perturbation independent "Lyapunov decay". \\

\newpage

Another recent observation \mbox{[Cohen PRE 2002]}
is in the context of dephasing due to the interaction
with chaotic degrees of freedom.
The same idea of having non-perturbative/semiclassical regime
is applicable. In fact the study of "dephasing" can be regarded
as a generalization of quantum irreversibility studies.
The calculation of the dephasing factor reduces to the study
of "fidelity" in a scenario where one has control over both the
amplitude $A$ and the rate $V$ of the driving. \\

\ \\

\noindent {\bf References:} \\

\noindent
For a pedagogical presentation, including references,
see D. Cohen, "Driven chaotic mesoscopic systems,dissipation and decoherence",
lecture notes of the course to be given in the 2002 Wroclaw school.
A preliminary version can be found in http://www.bgu.ac.il/$\sim$dcohen.

\end{document}